\documentclass[11pt,onecolumn,letterpaper]{IEEEtran}
\usepackage{amsmath,amstext, amsfonts, amssymb}

\newcommand{\ie}{{\em i.e.}}
\newcommand{\E}{\textsf{E}}
\newtheorem{lemma}{Lemma}[section]
\renewcommand{\vec}[1]{{\textbf{\textit{#1}}}}

\begin{document}
\title  {Variations on a theme by Schalkwijk and Kailath}
\author{Robert G. Gallager \quad  \and \quad  Bar\i\c{s} Nakibo\u{g}lu}
\date{\today}

\maketitle
\begin{abstract}
Schalkwijk and Kailath (1966) developed a class of block  codes for 
Gaussian channels with ideal feedback for which the probability of 
decoding error decreases as a second-order exponent in block length 
for rates below capacity.  This well-known but surprising result  is 
explained and simply  derived here in terms of a result by Elias 
(1956) concerning the minimum  mean-square distortion achievable in 
transmitting a single Gaussian random  variable over multiple uses of 
the same Gaussian channel.   A simple  modification of the 
Schalkwijk-Kailath scheme is then shown to have an error probability 
that decreases  with an  exponential \emph{order} which is linearly 
increasing with block length.  In the infinite bandwidth limit, this 
scheme produces zero error probability using bounded expected energy 
at all rates below capacity.  A lower  bound on error probability for 
the finite bandwidth case is then derived  in which the error 
probability decreases with an exponential  order which is linearly 
increasing in block length at the same rate as the upper bound.
\end{abstract}

\section {Introduction}\label{sec1}
This note describes coding and decoding strategies for discrete-time 
additive memoryless Gaussian-noise  (DAMGN) channels with ideal 
feedback.  It was shown by  Shannon \cite{Shan} in 1961 that feedback 
does not increase the capacity of memoryless channels, and was shown 
by Pinsker \cite{Pin} in 1968 that fixed-length block codes on 
Gaussian-noise channels with feedback can not   exceed the sphere 
packing bound if the energy per codeword is bounded independently of 
the noise realization. It is clear, however, that reliable 
communication  can be simplified by the use of feedback, as 
illustrated by standard automatic repeat strategies at the data link 
control layer.  There is a substantial literature (for example 
\cite{Sahai}, \cite{Burn}, \cite{NG}) on using variable-length 
strategies to substantially improve the rate of exponential decay of 
error probability with \emph{expected}  coding constraint length. 
These strategies essentially use the feedback to coordinate 
postponement of the final decision  when the noise would otherwise cause 
errors.  Thus small error probabilities can be achieved through the 
use of occasional long delays, while keeping the expected delay small.

  For DAMGN channels an additional mechanism for using feedback exists 
whereby  the transmitter can transmit unusually large amplitude 
signals   when it observes that  the receiver   is in danger of 
making a decoding error.  The power (\ie, the expected squared 
amplitude) can be kept small because these large amplitude signals 
are rarely required.  In 1966, Schalkwijk and Kailath \cite{SchalkK} 
used this mechanism in a fixed-length block-coding scheme for 
infinite bandwidth Gaussian noise channels with ideal feedback.  They 
demonstrated the surprising result that the resulting probability of 
decoding error decreases as a second order exponential\footnote{For 
integer $k\ge 1$, the $k$th order  exponent function $g_k(x)$ is 
defined as $g_k(x) = \exp(\exp(\cdots(\exp(x))\cdots))$ with $k$ 
repetitions of exp.   A function $f(x)\ge 0$ is said to decrease as a 
$k$th order exponential if for some constant $A>0$ and all 
sufficiently large $x$, $f(x) \le 1/g_k(Ax)$.}
 in the   code constraint length at all 
transmission rates less than capacity. Schalkwijk \cite{Schalk} 
extended this result to the finite bandwidth case, \ie, DAMGN 
channels.  Later, Kramer \cite{Kramer} (for the infinite bandwidth 
case) and Zigangirov \cite{Zig} (for the finite bandwidth case) 
showed that the above doubly exponential bounds could be replaced by 
$k$th order exponential bounds for any $k>2$ in the limit of 
arbitrarily large block lengths.  Later encoding schemes inspired by 
the Schalkwijk and Kailath approach have been developed for 
multi-user communication  with DAMGN \cite{Ozarow}, \cite{Ozarow2}, 
\cite{KramerJ}, \cite{Bross}, \cite{Stark}, secure communication with 
DAMGN \cite{Deniz} and point to point communication for Gaussian 
noise channels  with memory \cite{YHK}.

The purpose of this paper is three-fold.  First, the existing results 
for DAMGN  channels with ideal feedback are made more transparent by 
expressing them in terms of a 1956 paper by Elias on transmitting a 
single signal from a Gaussian source via multiple uses of a DAMGN 
channel with feedback.  Second, using an approach similar to that of 
Zigangirov in \cite{Zig}, we strengthen the results of \cite{Kramer} 
and \cite{Zig}, showing that error probability can be made to 
decrease with blocklength $n$ at least with an exponential order $a n 
-b$ for given coefficients $a>0$ and $b>0$.  Third, a lower bound is 
derived. This lower bound decreases with an exponential order in $n$ 
equal to $an +b'(n)$ where $a$ is the same as in the upper bound
and $b'(n)$ is a sublinear  function
\footnote{i.e. $\displaystyle{\lim_{n \rightarrow \infty} \tfrac{b'(n)}{n}=0}$}
 of the block length $n$.

Neither this paper nor the earlier results in \cite{Schalk}, 
\cite{SchalkK}, \cite{Kramer}, and \cite{Zig} are intended to be 
practical. Indeed, these second and higher order exponents require 
unbounded amplitudes (see \cite{Pin}, \cite{Burn1}, \cite{NG}).  Also 
Kim et al \cite{Kim} have recently shown that if the feedback is 
ideal except for additive Gaussian noise, then the error probability 
decreases only as a single exponential in block length, although the 
exponent increases with increasing signal-to-noise ratio in the 
feedback channel. Thus our purpose here is simply to provide 
increased  understanding of the ideal conditions assumed.

We first review the Elias result \cite{Elias} and use it to get an 
almost trivial derivation of the Schalkwijk and Kailath results. The 
derivation yields  an exact expression for error probability, 
optimized over a class of algorithms including those in 
\cite{Schalk},  \cite{SchalkK}.  The linear processing 
inherent in that class of algorithms is relaxed to obtain error 
probabilities that decrease with block length $n$ at a rate much faster than 
an exponential order of 2. Finally a lower bound  to the probability of 
decoding error is derived.  This lower bound is first derived for the 
case of two codewords and is then generalized to arbitrary rates less 
than capacity.

\section{The feedback channel and  the Elias result}
Let $X_1, \ldots, X_n = {\bf X}_1  ^{n}$ represent $n>1$ successive 
inputs to a discrete-time additive memoryless Gaussian noise (DAMGN) 
channel with ideal feedback.  That is, the channel outputs $Y_1, 
\ldots, Y_n = {\bf Y}_1^{n}$ satisfy ${\bf Y}_1^{n} = {\bf X}_1^{n} + 
{\bf Z}_1^{n}$  where ${\bf Z}_1^{n}$ is an $n$-tuple of 
statistically independent  Gaussian random variables, each with zero 
mean  and variance $\sigma^2_Z$,  denoted ${\mathcal N}(0, 
\sigma^2_Z)$.    The channel  inputs are constrained to some given 
average power constraint $S$ in the sense that the inputs must 
satisfy the second-moment constraint
\begin{equation}
\label{pc}
\frac{1}{n}\sum_{i=1}^n S_i\le S \qquad \qquad
\mbox{where}\,\,S_i = \E[X_i^2] .
\end{equation}
Without loss of generality, we take $\sigma_Z^2 = 1$. Thus $S$ is 
both a power constraint and a signal-to-noise ratio constraint.

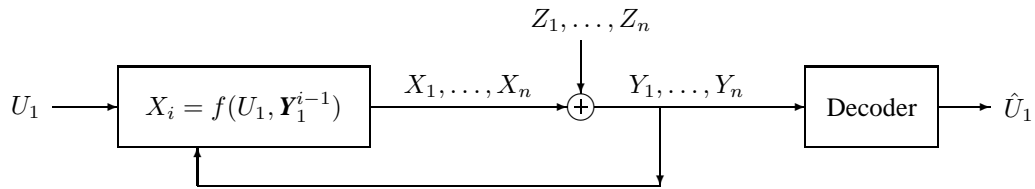
\begin{figure*}[t]
\centering
\setlength{\unitlength}{5pt}
\begin{picture}(64,16)(0,0)
{\small\put(-5,4){\makebox(8,6){$U_1$}}
\put(1,7){\vector(1,0){5}}
\put(6,4){\framebox(19,6){\small $X_i = f(U_1, \vec Y_1^{i-1})$}}
\put(25,7){\vector(1,0){15}}
\put(42,7){\vector(1,0){16}}
\put(45,5.5){\makebox(8,6){$Y_1, \ldots, Y_n$}}
\put(47, 7){\vector(0,-1){6}}
\put(47, 1){\line(-1, 0){35}}
\put(12, 1){\vector(0,1){3}}
\put(28.5,5.5){\makebox(8,6){$X_1, \ldots, X_n$}}
\put(41,7){\circle{2}}\put(41,12){\vector(0,-1){4}}
\put(41,7.5){\line(0,-1){1}}\put(40.5,7){\line(1,0){1}}
\put(58,4){\framebox(10,6){\small Decoder}}
\put(37,10.5){\makebox(9.5,6){$Z_1, \ldots, Z_n$}}
\put(73,4){\makebox(6,6)[l]{$   {\hat U_1}$}}
\put(68,7){\vector(1,0){4}}}
\end{picture}
\vspace{-.3cm}
\caption{\footnotesize   The setup for $n$ channel uses per source use with ideal feedback. }
\label{figElias}
\vspace{-0.2cm}
\end{figure*}

A discrete-time channel is said to have ideal feedback if  each 
output $Y_i,\, 1\le i \le n$, is made known to the transmitter in 
time to generate input $X_{i+1}$ (see Figure  \ref{figElias}).  Let 
$U_1$ be the random source  symbol to be communicated via this 
$n$-tuple of channel uses.  Then each channel   input $X_i$ is some function 
$f(U_1, {\bf Y}_1^{i-1})$ of the source and previous outputs.  Assume 
(as usual) that $U_1$ is statistically independent of ${\bf Z}^{n}_1$.

Elias \cite{Elias} was interested in the situation where $U_1\sim 
\mathcal N(0, \sigma_1^2)$ is a Gaussian random variable rather than 
a discrete message.  For $n=1$, the rate-distortion bound (with a 
mean-square distortion measure) is achieved without coding or 
feedback.   For $n>1$, attempts to map $U_1$ into an $n$ dimensional 
channel input in the absence of feedback involve non-linear or 
twisted modulation techniques that are ugly at best. Using the ideal 
feedback, however, Elias constructed a simple and elegant procedure 
for using the $n$ channel symbols to send $U_1$ in such a way as to 
meet the rate-distortion bound with equality.

Let $S_i = \E[X_i^2]$  be an arbitrary choice of energy, \ie, second 
moment,  for each $i$,  $1\le i \le n$. It will be shown shortly that 
the optimal choice for $S_1, \ldots, S_n$, subject to (\ref{pc}), is 
$S_i = S$ for $1\le i \le n$.   Elias's strategy starts by choosing 
the first transmitted signal $X_1$ to be a linear scaling of the 
source variable $U_1$, scaled to meet the second-moment constraint, 
\ie, 
\begin{equation*}
  X_1 = \tfrac{\sqrt{S_1}U_1}{\sigma_1}.
\end{equation*}
   At the receiver, the 
minimum mean-square error (MMSE) estimate  of $X_1$ is $\E[X_1|Y_1] = 
\frac{S_1 Y_1}{1+S_1}$,  and the error in that estimate is $\mathcal 
N(0,\, \frac{S_1}{1+S_1})$. It is more convenient to keep track of 
the MMSE estimate of $U_1$ and the error $U_2$ in 
that estimate. Since $U_1$ and $X_1$ are the same except for the 
scale factor $\sigma_1/\sqrt{S_1}$, these are given by
\begin{align}
\label{estimate1a}
\E[U_1|Y_1]  &= \frac{\sigma_1\sqrt{S_1}\,Y_1}{1+ S_1}\\
\label{estimate1b}
U_2&= U_1 - \E[U_1|Y_1]  
\end{align}
where $U_2\sim \mathcal N(0, \sigma_2^2)$ and $\sigma^2_2 = \tfrac{\sigma^2_1}{1+ S_1}$.

Using the feedback, the transmitter can calculate the error term 
$U_2$ at  time $2$.  Elias's strategy is to use $U_2$ as the source 
signal (without a second-moment constraint) for the second 
transmission.  This unconstrained signal $U_2$ is then linearly 
scaled to meet the second moment constraint $S_2$ for the second 
transmission. Thus the second transmitted signal $X_2$ is given by
\begin{equation*}
 X_2 = \tfrac{\sqrt{S_2}U_2}{\sigma_2}. 
\end{equation*}
  We use this notational device 
throughout, referring to the unconstrained source signal to be sent 
at time $i$ by $U_i$ and to the linear scaling of $U_i$, scaled to 
meet the second moment constraint $S_i$, as $X_i$.

  The receiver calculates the MMSE estimate $\E[U_2|Y_2] = 
\frac{\sigma_2\sqrt{S_2}\,Y_2}{1+ S_2}$ and the transmitter then 
calculates the error in this estimate, $U_3 = U_2 - \E[U_2|Y_2]$. 
Note that
\begin{align*}
  U_1 
&= U_2+\E[U_1|Y_1]\\
&= U_3 + \E[U_2|Y_2] + \E[U_1|Y_1] .
\end{align*}
Thus $U_3$ can be viewed as the error arising from estimating $U_1$ 
by $\E[U_1|Y_1] + \E[U_2|Y_2]$.  The receiver continues to update its 
estimate of $U_1$ on subsequent channel uses, and the transmitter 
continues to transmit linearly   scaled versions of the current 
estimation error.  Then the general expressions  are as follows:
\begin{align}
\label{estimateia}
X_i
&= \frac{\sqrt{S_i}U_i}{\sigma_i}\,;\\
%\qquad\quad 
\label{estimateib}
\E[U_i|Y_i]  
&=  \frac{\sigma_i\sqrt{S_i}\,Y_i}{1+ S_i};\\
\label{estimateic}
U_{i+1} 
&=  U_i - \E[U_i|Y_i].  
\end{align}
where 
$U_{i+1} \sim \mathcal N(0, \sigma_{i+1}^2)$
and 
$\sigma_{i+1}^2 = \frac{\sigma_i^2}{1+S_i}$.

Iterating on equation (\ref{estimateic})  from $i = 1$ to $n$ yields
\begin{equation}
    \label{finala}
U_{n+1} = U_1 -\sum_{i=1}^{n} \E[U_i|Y_i]. 
\end{equation}
Similarly, iterating on $\sigma_{i+1}^2 =\sigma_i^2/(1+S_i)$, we get
\begin{equation}
\label{finalb}
\sigma_{n+1}^2 
= \frac{\sigma_1^2}{\prod_{i=1}^{n} (1+S_i)}.
\end{equation}

This says that the error arising from estimating $U_1$ by 
$\sum_{i=1}^{n} \E[U_i|Y_i]$  is $\mathcal N(0, \sigma_{n+1}^2)$. 
This is valid for any (non-negative) choice of $S_1, \ldots, S_n$, 
and this is minimized, subject to $\sum_{i=1}^{n} S_i = nS$, by $S_i 
= S$ for $1\le i \le n$. With this optimal assignment, the mean 
square estimation error in  $U_1$ after $n$ channel uses is
\begin{equation}
\label{iterV1}
\sigma_{n+1}^2 = \frac{\sigma_1^2}{{(1+S)^{n}}}\,.
\end{equation}
We now show that this is the minimum mean-square error over all ways 
of using the channel.  The rate-distortion function for this Gaussian 
source with a squared-difference distortion measure  is well known to 
be
\begin{equation*}
R(d) = \frac{1}{2} \ln \frac{\sigma_1^2}{d}
\end{equation*}
This is the minimum mutual information, over all channels, required 
to achieve a mean-square error (distortion) equal to $d$.  For $d = 
\sigma_1^2/(1+S)^{n}$, $R(d)$ is $\frac{n}{2} \ln (1+S)$, which is 
the capacity of this channel over $n$ uses (it was shown by Shannon 
\cite{Shan} that feedback does not increase the capacity of 
memoryless channels).  Thus the Elias scheme actually meets the 
rate-distortion bound with equality, and no other coding system, no 
matter how complex, can achieve a smaller mean-square error.  Note 
that (\ref{iterV1}) is also valid in the degenerate case $n=1$. What 
is surprising about this result is  not so much that it meets the 
rate-distortion bound, but rather that the mean-square estimation 
error goes down geometrically with $n$.  It is this property that 
leads directly to the doubly exponential error probability of the 
Schalkwijk-Kailath scheme.

\section{The Schalkwijk-Kailath scheme}\label{secSK}
The Schalkwijk and Kailath (SK) scheme will now be defined in terms 
of the Elias  scheme,\footnote{The analysis here is tutorial and was 
carried out in slightly simplified form in \cite[p481]{Gall}.  
A very readable further simplified analysis is in \cite{KimT}.}
 still assuming the discrete-time channel model 
of Figure \ref{figElias} and the power constraint of (\ref{pc}). The 
source is a set of $M$ equiprobable symbols, denoted by $\{1, 2, 
\ldots,  M\}$. The channel uses will now be numbered from $0$ to 
$n-1$, since the use at time 0 will be quite distinct from the 
others.  The source signal, $U_0$  is a standard $M$-PAM modulation 
of the source symbol.     That is, for each symbol $m$, $1\le m \le 
M$, from the source alphabet, $m$ is mapped into the signal $a_m$ 
where $a_m = m-(M{+}1)/2$.  Thus the $M$ signals in $U_0$ are 
symmetric around 0 with unit spacing.      Assuming equiprobable 
symbols, the second moment $\sigma_0^2$ of $U_0$ is  $(M^2-1)/12$. 
The initial channel input $X_0$ is a linear scaling of $U_0$, scaled 
to have an energy $S_0$ to be determined later.  Thus $X_0$ is an 
$M$-PAM encoding, with signal separation $d_0 = \sqrt{S_0}/\sigma_0$.
\begin{equation}
\label{alph}
X_0 \,=\, U_0 \sqrt{\frac{S_0}{\sigma^2_0}} \,= \,U_0 
\sqrt{\frac{S_0}{12(M^2-1)}} .
\end{equation}
The received signal  $Y_0 = X_0+Z_0$ is fed back to the transmitter, 
which, knowing $X_0$, determines $Z_0$.  In the following $n-1$ 
channel uses,  the Elias scheme is used to send the Gaussian random 
variable $Z_0$ to the receiver, thus reducing the effect of the noise 
on the original transmission. After the $n-1$ transmissions to convey 
$Z_0$, the receiver combines its estimate of $Z_0$ with $Y_0$ to get 
an estimate of $X_0$, from which the $M$-ary signal is detected.

Specifically, the transmitted and received signals for times $1\le i 
\le n-1$ are given by equations (\ref{estimateia}), (\ref{estimateib})  and
(\ref{estimateic}).  At time 
1,  the unconstrained signal $U_1$ is $Z_0$ and $\sigma^2_1 = 
\E[U_1^2] = 1$.  Thus the transmitted signal $X_1$ is given by 
$\sqrt{S_1} U_1$, where the second moment $S_1$ is to be selected 
later. We choose $S_i = S_1$ for $1\le i \le n-1$ for optimized use 
of the Elias scheme, and thus the power constraint in  (\ref{pc}) 
becomes $S_0 +(n-1)S_1 = nS$.  At the end of transmission $n-1$,  the 
receiver's estimate of $Z_0$ from $Y_1, \ldots, Y_{n-1}$ is given by 
(\ref{finala}) as
\begin{equation*}
   \E[Z_0 \mid {\bf Y}_1^{n-1}] = \sum_{i=1}^{n-1} \E[U_i\mid Y_i].
\end{equation*}
The error in this estimate, $U_n = Z_0 - \E[Z_0 \mid {\bf 
Y}_1^{n-1}]$, is a zero-mean Gaussian random variable with variance 
$\sigma^2_n$, where $\sigma_n^2$ is given by  (\ref{iterV1}) to be
\begin{equation}
\label{msen}
\sigma^2_n =  \frac{1}{(1+S_1)^{n-1}}.
\end{equation}
Since $ Y_0=X_0+ Z_0$ and $Z_0 = \E[Z_0\mid \vec Y_1^{n-1}] + U_n$ we have
\begin{equation}
\label{geonoise}
Y_0 - \E[Z_0 \mid {\bf Y}_1^{n-1}] = X_0 + U_n 
\end{equation}
where $U_n \sim \mathcal N(0, \sigma^2_n).$
  
Note that $U_n \sim \mathcal N(0, \sigma^2_n)$ is a function of the 
noise vector $\vec Z_0^{n-1}$ and is thus statistically 
independent\footnote{Furthermore, for the given feedback strategy, 
Gaussian estimation theory can be used to show, first, that $U_n$ is 
independent of  $\E[Z_0 \mid {\bf Y}_1^{n-1}]$, and, second, that 
$\tilde Y = Y_0 - \E[Z_0 \mid {\bf Y}_1^{n-1}]$ is a sufficient 
statistic for $X_0$ based on $\vec Y_0^{n-1}$, (i.e.  $\Pr[X_0\mid 
{\bf Y}_0^{n-1}]=\Pr[X_0\mid \tilde{Y}]$). Thus this detection 
strategy is not as ad hoc as it might initially seem. } of $X_0$. 
Thus, detecting $X_0$ from $Y_0 - \E[Z_0 \mid {\bf Y}_1^{n-1}]$ 
(which is known at the receiver.) is the simplest of classical 
detection problems, namely that of detecting an $M$-PAM signal $X_0$ 
from the signal plus an independent Gaussian noise variable $U_n$. 
Using maximum likelihood detection, an error occurs only if $U_n$ 
exceeds half the distance between signal points, \ie, if  $|U_n| \ge 
\frac{1}{2}\,\frac{\sqrt{S_0}}{\sigma_0} = 
\frac{1}{2}\,\sqrt{\frac{12S_0}{M^2-1}}$.  Since the variance of 
$U_n$ is $(1+S_1)^{-n+1}$, the probability of error is given 
by\footnote{The term$(M{-}1)/M$ in (\ref{Pe1}) arises because the 
largest and smallest signals each have only one nearest neighbor, 
whereas all other signals have two nearest neighbors.}
\begin{equation}
\label{Pe1}
P_e = 2\frac{(M{-}1)}{M} Q(\gamma_n)
\end{equation}
where ${ \gamma_n = \tfrac{1}{2}\sqrt{\frac{12S_0(1+S_1)^{n-1}}{M^2-1}}}$ and $Q(x)$ is the complementary  distribution function of $\mathcal N(0,1)$,  \ie,
\begin{equation}
\label{eq:qf}
Q(x)=\frac{1}{\sqrt{2\pi}} \int_x^\infty \exp(\frac{-z^2}{2})\,dz.
\end{equation}
Choosing $S_0$ and $S_1$, subject $S_0 + (n{-}1)S_1 = nS$, to 
maximize $\gamma_n$ (and thus minimize $P_e$), we get $S_1 = \max\{0, 
S - \frac{1}{n}\}$. That is, if $nS$ is less than 1, all the energy 
is used to send $X_0$ and the feedback is unused.  We assume $nS>1$ 
in what follows,  since for any given $S>0$ this holds for large 
enough $n$. In this case, $S_0$ is one unit larger than $S_1$, 
leading to
\begin{equation}
\label{enerdis}
S_1 = S - \frac{1}{n};\qquad S_0 = S_1+1.
\end{equation}
Substituting (\ref{enerdis}) into (\ref{Pe1}),
\begin{equation}
\label{Pe3}
P_e = 2\frac{(M{-}1)}{M} Q(\gamma_n)
\end{equation}
where $\gamma_n = \sqrt{\frac{3(1+S - \frac{1}{n})^{n}} {M^2-1}}.$

This is an \emph{exact} expression for error probability, optimized 
over  energy distribution, and using $M$-PAM followed by the Elias 
scheme and ML detection. It can be simplified as an upper bound by 
replacing the coefficient $\tfrac{M-1}{M}$ by 1.  Also, since 
$Q(\cdot)$ is a decreasing function of its argument, $P_e$ can be 
further upper bounded by replacing $M^2-1$ by $M^2$. Thus,
\begin{equation}
\label{Pe3a}
P_e \le 2Q(\gamma_n)
\end{equation}
where $\gamma_n \ge \sqrt{3} \left(1- \frac{1}{(1+S)n}\right)^{n/2} 
\frac{(1+S)^{n/2}}{M}.$

For large $M$, which is the case of interest, the above bound is very 
tight and is essentially an equality, as first derived by 
Schalkwijk\footnote{Schalkwijk's work was independent of Elias's.  He 
interpreted the steps in the algorithm as successive improvements in 
estimating $X_0$ rather than as estimating $Z_0$.} in Eq. 12 of 
\cite{Schalk}. Recalling that $nS\ge 1$ we can further lower bound 
$\gamma_n$ (thus upper bounding $P_e$). Substituting 
$C(S)=\frac{1}{2}\ln (1+S)$ and $M=\exp(nR)$ we get
\begin{equation}
\label{Pe4}
\gamma_n \ge  \left[\sqrt{3} (1-\frac{1}{1+n})^{n/2}\right] \exp(n(C(S)-R))
\end{equation}
The term in brackets is decreasing in $n$. Thus,
\begin{align}
(1-\tfrac{1}{1+n})^{n/2}
&\geq \lim_{k \rightarrow \infty}   (1-\tfrac{1}{1+k})^{k/2}&~&~\\
&\geq e^{-1/2}  &~& \forall n\geq 1
\end{align}
Using this together with equations (\ref{Pe3a}) and  (\ref{Pe4}) we get,
\begin{eqnarray}
\label{Pe4a}
P_e \le 2Q\left(\sqrt{\tfrac{3}{e}} \exp(n (C(S)-R)) \right),
\end{eqnarray}
or more simply yet,
\begin{equation}
\label{Pe4aa}
P_e \le 2Q(\exp[n(C(S)-R)]).
\end{equation}
Note that for $R<C(S)$,  $P_e$ decreases as a second order exponential in $n$.

  In summary, then, we see that the use of standard $M$-PAM at time 0, 
followed by  the Elias algorithm  over  the next $n-1$ transmissions, 
followed by ML detection, gives rise to a probability of error $P_e$ 
that  decreases as a second-order exponential for all $R<C(S)$. Also 
$P_e$ satisfies (\ref{Pe4a}) and (\ref{Pe4aa}) for all $n\ge 1/S$.

Although $P_e$ decreases as a second-order exponential with this 
algorithm, the algorithm does not minimize $P_e$ over all algorithms 
using ideal feedback.  The use of standard $M$-PAM at time 0 could be 
replaced by PAM with non-equal spacing of the signal points for a 
modest reduction in $P_e$.  Also, as shown in the next section, 
allowing transmissions  1 to $n-1$ to make use of the discrete nature 
of $X_0$ allows for a major reduction in $P_e$.\footnote{Indeed, 
Zigangirov \cite{Zig} developed an algorithm quite similar to that 
developed in the next section.  The initial phase of that algorithm 
is very similar to the algorithm \cite{Schalk} just described, with 
the following differences.  Instead of starting with standard 
$M$-PAM, \cite{Zig} starts with a random ensemble of 
non-equally-spaced $M$-PAM codes ingeniously arranged to form a 
Gaussian random variable.  The Elias scheme is then used, starting 
with this Gaussian random variable. Thus the algorithm in \cite{Zig} 
has different  constraints than those above.  It turns out to have an 
insignificantly larger $P_e$ (over this phase) than the algorithm 
here for $S$ greater than $[(1/\ln \tfrac{6}{\pi})-1]$ and an 
insignificantly smaller $P_e$ otherwise. }

  The algorithm above, however, does have the property that it is 
optimal  among schemes in which, first, standard PAM is used at time 
0 and, second, for each $i$, $1\le i \le n-1$, $X_i$  is a linear 
function of $Z_0$ and $Y_1^{i-1}$. The reason for this is that $Z_0$ 
and $Y_1^{n-1}$ are then jointly Gaussian and the Elias scheme 
minimizes the mean square error in $Z_0$ and thus also minimizes 
$P_e$.

\subsection{Broadband Analysis:}

Translating these results to a continuous time formulation where the 
channel is used 2W times per second,\footnote{This is usually referred 
to as a channel bandlimited to $W$.  This is a harmless and 
universally used abuse of the word bandwidth for channels without 
feedback, and refers to the ability to satisfy the Nyquist criterion 
with arbitrarily little power sent out of band.  It is more 
problematic with feedback, since it assumes that the sum of the 
propagation delay, the duration of the transmit pulse, the duration 
of the matched filter at the receiver, and the corresponding 
quantities for the feedback, is at most $1/2W$.  Even allowing for a 
small fraction of out-of-band energy, this requires considerably more 
than bandwidth $W$.}  the capacity (in nats per second) is $C_W = 
2WC$.  Letting $T = n/2W$ and letting $R_W = 2WR$ be the rate in nats 
per second, this formula becomes
\begin{equation}
\label{Pe5}
P_{e} \le 2\,Q\left(\exp \left[(C_W - R_W)T  \right] \right).
\end{equation}
Let $\mathcal P = 2WS$ be the continuous-time power constraint, so 
that $C_W = W\ln (1+\mathcal P/2W)$.  In the broadband limit as $W 
\to \infty$ for fixed $\mathcal P$, $C_W \to \mathcal P/2$.   Since 
(\ref{Pe5}) applies for all $W>0$, we can  simply go to the broadband 
limit,  $C_\infty = \mathcal P/2$.   Since the algorithm is basically 
a discrete time algorithm, however, it  makes more sense to view the 
infinite bandwidth limit as a limit in which the number of available 
degrees of freedom $n$ increases faster than linearly with the 
constraint time $T$.  In this case, the signal-to-noise ratio per 
degree of freedom, $S = \mathcal PT/n$ goes to 0 with increasing $T$. 
Rewriting  $\gamma_n$ in (\ref{Pe3a}) for this case,
\begin{eqnarray}
\label{Pe6}
\gamma_n &\ge&\sqrt{3}\exp\left[\frac{n}{2}\ln(1+\frac{\mathcal 
PT}{n} - \frac{1}{n})-T R_{\infty}\right]\\
&\ge& \sqrt{3}\exp\left[\frac{\mathcal P T}{2} - \frac{1}{2} - 
\frac{\mathcal P^2T^2}{4n}-TR_\infty\right],
\label{bign}
\end{eqnarray}
where the inequality $\ln(1+x) \ge x-x^2/2$ was used.  Note that if 
$n$ increases quadratically with $T$, then the term $\frac{\mathcal 
P^2T^2}{4n}$ is simply a constant which becomes negligible as the 
coefficient on the quadratic becomes large.  For example, if $n \ge 
6\mathcal P^2 T^2$, then this  term is at most $1/24$ and 
(\ref{bign}) simplifies to
\begin{equation}
\gamma_n \ge  \,\exp   \left[T(C_\infty-R_\infty) \right]
\qquad \mbox{for}\quad n \ge 6\mathcal P^2T^2.
\label{Pe8}
\end{equation}
This is  essentially the same as the broadband  SK result (see the 
final equation in  \cite{SchalkK}).  The result in \cite{SchalkK} 
used $n = e^{2TC_W}$ degrees of freedom, but chose the subsequent 
energy levels to be decreasing harmonically, thus slightly weakening 
the coefficient of the result. The broadband result is quite 
insensitive to the energy levels used for each degree of 
freedom\footnote{To see this, replace $(1+S_1)^{(n-1)/2}$ in 
(\ref{Pe1}) by $\frac{1}{2} \exp[\,\sum_i\ln(1+S_i)]$, each term of 
which can be lower bounded by the inequality $\ln (1+x) \ge 
x-x^2/2$.}, so long as $S_0$ is close to 1 and the other $S_i$ are 
close to 0.  This partly explains why the harmonic choice of energy 
levels in \cite{SchalkK} comes reasonably close to the optimum result.

\section{An alternative PAM Scheme in the high signal-to-noise
regime}\label{secPAM}
In the previous section, Elias's scheme was used to allow the receiver
to estimate the noise $Z_0$ originally added to the PAM signal at time $0$.
This gave rise  to an equivalent observation, $Y_0 - \E[Z_0 \mid
{\bf Y}_1^{n-1}]$  with  attenuated noise $U_n$ as given in
(\ref{geonoise}).  The geometric attenuation of $\E[U_n^2]$ with $n$
  is the reason why the error probability
in the Schalkwijk and Kailath (SK) \cite{SchalkK} scheme decreases
as a second order exponential in time.

\begin{figure*}[t]
\setlength{\unitlength}{6pt}
\centering
\begin{picture}(76,18)(-35,-5)
\put(-35,0){\line(1,0){80}}
\put(0,0){\vector(0,1){11.5}}
\put(28,0){\vector(0,1){11.5}}
\put(-28,0){\vector(0,1){11.5}}
\put(14,0){\line(0,1){5}}
\put(-14,0){\line(0,1){5}}
\put(42,0){\line(0,1){5}}
\qbezier(0,10)(2.1078,10)(6,6.065)
\qbezier(6,6.065)(9.582,2.444)(12,1.353)
\qbezier(12,1.353)(14.292,.317)(19.2,.111)
\qbezier(0,10)(-2.1078,10)(-6,6.065)
\qbezier(-6,6.065)(-9.582,2.444)(-12,1.353)
\qbezier(-12,1.353)(-14.292,.317)(-19.2,.111)
\put(-2,-4.5){\makebox(4,4)[l]{\footnotesize $a_m d_0$}}
\put(26,-4.5){\makebox(4,4)[l]{\footnotesize $a_{m{+}1}
d_0$}}
\put(-31,-4.5){\makebox(4,4)[l]{\footnotesize $a_{m{-}1}
d_0$}}
\put(4.7,7.0){\makebox(4,4)[l]{\footnotesize $f(Y_0\mid U_0=a_m)$}}
\put(2.5,0.2){\makebox(4,4)[l]{\footnotesize $\hat m = m$}}
\put(31,1.0){\makebox(4,4)[l]{\footnotesize $\hat m = m{+}1$}}
\put(0,3){\vector(1,0){14}}
\put(0,3){\vector(-1,0){14}}
\put(28,2){\vector(1,0){14}}
\put(28,2){\vector(-1,0){14}}
\end{picture}\vspace{-.3cm}
\caption{ \footnotesize Given that $a_m$ is the sample value of the 
PAM source signal $U_0$, the sample value of  $X_0$ is $a_m d_0$ where 
$d_0 = \sqrt{S_0}/\sigma_0$.  The figure illustrates the probability 
density of $Y_0$ given this conditioning and shows  the $M$-PAM 
signal points for $X_0$ that are neighbors to the sample value $X_0 
=a_m d_0$.   Note that this density is
$\mathcal N(a_m d_0,\, 1)$, \ie, it is the density of $Z_0$, shifted 
to be centered at $a_m d_0$.
Detection using maximum likelihood at this point simply quantizes
$Y_0$ to the nearest signal point.}
\label{Det}
\vspace{-0.2cm}
\end{figure*}
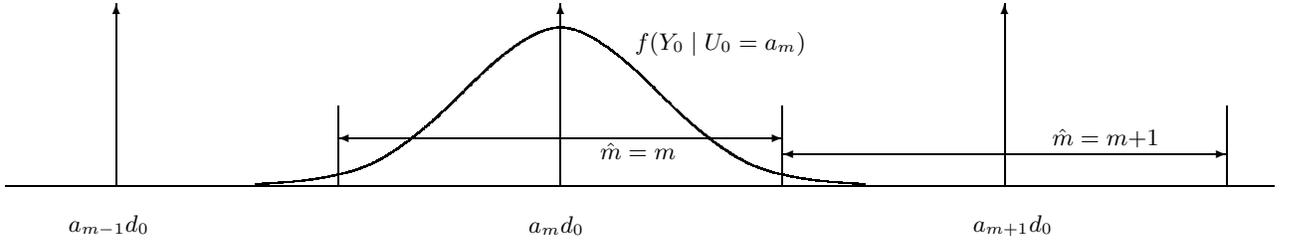

In this section, we explore an alternative strategy that is again based
on the use of $M$-PAM  at time 0, but is quite different from the SK 
strategy at times 1 to $n-1$.  The analysis is restricted to
situations in which the signal-to-noise ratio (SNR) at time 0 is so 
large that the distance between successive PAM  signal points in 
$X_0$ is large relative to the standard deviation of the noise.  In 
this high SNR regime, a simpler and more effective strategy than the 
Elias scheme suggests itself  (see Figure \ref{Det}).  This new 
strategy is limited to the high SNR regime, but Section 
\ref{sectwophase} develops a two-phase scheme that
uses the SK strategy for the first part of the block, and switches to this
new strategy when the SNR is sufficiently large.

In this new strategy for the high SNR regime, the receiver makes a tentative
ML decision $\hat m_0$ at time 0. As seen in the figure,  that 
decision is  correct unless the noise exceeds half the distance $d_0 
= \sqrt{S_0}/\sigma_0$
to either the signal value on the right or the left of the sample value
$a_m$ of $U_0$.  Each of these two events  has probability   $Q(d_0/2)$.

The transmitter uses the feedback to calculate $\hat m_0$
and  chooses the next  signal $U_1$
(in the absence of  a second-moment constraint) to be
a shifted version of the original  $M$-PAM signal,  shifted so
that $U_1 = \hat m_0 - m$ where $m$ is the original message symbol being
transmitted. In other words, $U_1$ is the integer-valued error in the 
receiver's tentative decision $a_{\hat m_0}$ of $U_0$.  The 
corresponding transmitted signal
$X_1$ is essentially given by $X_1  = U_1\sqrt{S_1/\E[U_1^2]}$, where 
$S_1$ is  the energy allocated to $X_1$.

We now give an approximate explanation of why this strategy makes sense
and how the subsequent transmissions are chosen.  This is followed
by a  precise analysis. Temporarily ignoring the case where either $m=1$
or $m = M$ (\ie, where $a_m$ has only one neighbor),  $U_1$ is $0$
with probability $1- 2Q(d_0/2)$.    The probability
that $|U_1|$ is  two or more is essentially negligible, so
$U_1 = \pm 1$  with a probability approximately equal to  $2Q(d_0/2)$.
Thus
\begin{equation}\label{approxU}
E[U_1^2] \approx 2Q(d_0/2);\qquad \quad X_1 \approx 
\frac{U_1\sqrt{S_1}}{\sqrt{2Q(d_0/2})}
\end{equation}
This means that $X_1$ is not only a shifted version of $X_0$, but 
(since $d_0 = \sqrt{S_0}/\sigma_0$) is also scaled up by a
factor that is exponential in $S_0$ when $S_0$ is sufficiently large.  Thus
the separation between adjacent signal points in $X_1$ is 
exponentially increasing with $S_0$.

This also means that when $X_1$ is transmitted, the situation is 
roughly the same as that in Figure \ref{Det}, except that the 
distance between signal points is increased by a factor exponential 
in $S_0$.  Thus a tentative decision at time 1 will have an  error 
probability that decreases as a second order exponential in $S_0$.

Repeating the same procedure at time 2 will then give rise to a third
order exponential in $S_0$, etc.  We now turn to a precise analysis and
  description of the algorithm at times 1 to $n-1$.

The following lemma provides an upper bound to the second moment of 
$U_1$, which
was approximated in (\ref{approxU}).
\begin{lemma}
\label{lem:pam1}
For any $d \ge 4$, let $U$ be a $d$-quantization of a normal random
variable $Z \sim \mathcal N(0,1)$ in the sense that for each integer
$\ell$,  if $Z \in (d\ell-\frac{d}{2}, d\ell+\frac{d}{2}]$, then $U = 
\ell$.  Then
$\E[U^2]$ is upper bounded by
\begin{equation}\label{Ubound}
   \E[U^2 ]\leq \tfrac{1.6}{d} \exp [-\tfrac{d^2}{8}]
\end{equation}

\end{lemma}

Note from Figure \ref{Det} that, aside from a slight exception 
described below,  $U_1 = \hat m_0 - m$ is the same as the 
$d_0$-quantization of $Z_0$ where $d_0 = \sqrt{S_0}/\sigma_0$.  The 
slight exception is that $\hat m_0$ should always lie between $1$ and 
$M$.   If $Z_0 > (M-m+1/2)$, then $U_1 = M-m$, whereas the 
$d_0$-quantization takes on a larger integer value.  There is a 
similar limit for $Z_0 < 1-m -1/2$.  This reduces the magnitude of 
$U_1$ in the above exceptional cases, and thus reduces the second 
moment.  Thus the bound in the lemma  also applies to $U_1$. For 
simplicity in what follows, we avoid this complication by assuming 
that the receiver allows $\hat m_0$ to be larger
than $M$ or smaller than 1. This increases both the error probability and
the energy over true ML tentative decisions, so the bounds also apply 
to the case
with true ML tentative decisions.

\begin{proof}  From the definition of $U$,  we see that $U=\ell$ if
$Z \in (d\ell-\frac{d}{2}, d\ell+\frac{d}{2}]$. Thus, for $\ell\ge 1$,
\[\Pr[U=\ell] = Q(d\ell-\frac{d}{2}) - Q(d\ell+\frac{d}{2})\]
 From symmetry, $\Pr[U=-\ell] = \Pr[U=\ell]$, so the second moment of 
$U$ is given by
\begin{align*}
\E[U^2]
&= 2\sum_{\ell=1}^\infty \ell^2\left[ Q(d\ell-\frac{d}{2}) - 
Q(d\ell+\frac{d}{2})\right]\\
&=2 Q(d/2) + 2\sum_{\ell=2}^\infty [\ell^2 - (\ell-1)^2] 
\left[Q(d\ell-\frac{d}{2})\right].
\end{align*}
Using the standard upper  bound $Q(x)\leq 
\frac{1}{\sqrt{2\pi}\,x}\,\exp[-x^2/2]$ for $x>0$, and recognizing 
that   $\ell^2 - (\ell-1)^2 = 2\ell-1$, this becomes
\begin{align}
  {\nonumber}
\E[U^2]
&\leq \frac{4}{\sqrt{2\pi}\,d}\left\{\exp[-d^2/8] + \sum_{\ell=2}^\infty \exp[-(2\ell-1)^2d^2/8]\right\} \\
{\nonumber}
&= \frac{4}{\sqrt{2\pi}\,d}\exp[-d^2/8]\left\{1 + \sum_{\ell=2}^\infty \exp[-4\ell(\ell-1)d^2/8]\right\}\\
{\nonumber}
&\leq \frac{4}{\sqrt{2\pi}\,d}\exp[-d^2/8]\left\{\frac{1}{1-\exp(-d^2)}\right\}\\
&\leq \frac{1.6}{d} \exp[-\tfrac{d^2}{8}]
\qquad\mbox{for}\,\, d\ge 4.
   \label{gamN}
\end{align}
\end{proof}

We now define the rest of this new algorithm. We have defined the
unconstrained signal $U_1$ at time 1 to be $\hat m_0 - m$ but have not
specified the energy constraint to be used in amplifying $U_1$ to $X_1$.
The analysis is simplified by defining $X_1$ in terms of a specified 
scaling factor between $U_1$ and $X_1$.  The energy in $X_1$ is 
determined later by this scaling.  In particular, let
\begin{equation*}
X_1 = d_1 U_1\qquad \quad \mbox{where}\quad d_1 = 
\sqrt{8}\exp\left(\frac{d_0^2}{16}\right).
\end{equation*}
  The peculiar expression for $d_1$ above looks less peculiar when 
expressed  as $d_1^2/8 = \exp(d_0^2/8)$.  When $Y_1 = X_1 + Z_1$ is 
received, we can visualize the situation from Figure \ref{Det} again, 
where now $d_0$ is replaced by $d_1$.  The signal set for $X_1$ is 
again a PAM set but it now has  signal spacing $d_1$ and  is centered 
on the signal corresponding to the transmitted source symbol $m$. 
The  signals are no longer equally likely, but the analysis is 
simplified if a maximum likelihood tentative decision $\hat m_1$ is 
again made.  We see that $\hat m_1 = \hat m_0 - \hat Y_1$ where $\hat 
Y_1$ is the $d_1$-quantization of $Y_1$ (and where the receiver again 
allows $\hat m_1$ to be an arbitrary integer) .  We can now state the 
algorithm for each time $i$, $1 \le i \le n-1$.
\begin{eqnarray}
\label{newalg0}
d_i &=& \sqrt{8} \exp\left(\tfrac{d_{i-1}^2}{16}\right)\\
\label{newalg1}
X_i &=& d_i U_i \\
\label{newalg2} 
\hat m_i &=& \hat m_{i-1} - \hat Y_i \\
\label{newalg3}
U_{i+1} &=& \hat m_{i} - m.
\end{eqnarray}
where $\hat Y_i$ is the  $d_i$-quantization of $Y_i$.

\begin{lemma}
\label{lem:HighSNR}
For $d_0 \ge 4$, the algorithm of  (\ref{newalg0})-(\ref{newalg3}) 
satisfies the 
following for all alphabet sizes $M$ and all message symbols $m$:
\begin{eqnarray}\label{newalg4}
\frac{d_i^2}{8} &=& g_i(\frac{d_0^2}{8})\,\, \ge\,\, g_i(2). \\
  \label{newalg5}
\E[X_i^2] &\le& \frac{12.8}{d_{i-1}}.
\\ \label{newalg6} \sum_{i=1}^\infty \E[X_i^2] &\le& 5.
\\ \label{newalg7} \Pr(\hat m_i \ne m) &\le& 1/g_{i+1}(2),
\end{eqnarray}
where  $g_i(x) = \exp(\cdots(\exp(x))\cdots)$ with $i$ exponentials.
\end{lemma}

\begin{proof}  From the definition of $d_i$ in (\ref{newalg0}),
\[\frac{d_i^2}{8} = \exp(\frac{d_{i-1}^2}{8}) = 
\exp(\exp(\frac{d_{i-2}^2}{8})) = \cdots = g_i(\frac{d_0^2}{8})\]
This establishes the first part of (\ref{newalg4}) and the inequality 
follows since $d_0 \ge 4$ and $g_i(x)$ is increasing in $x$.

Next, since $X_i = d_i U_i$, we can use (\ref{newalg4}) and Lemma 
\ref{lem:pam1} to see that
\begin{align*}
  \E[X_i^2] 
&= d_i^2 \E[U_i^2] \\
&= \left(8 \exp(\frac{d_{i-1}^2}{8})\right)\left(
\frac{1.6}{d_{i-1}} \exp(-\frac{d_{i-1}^2}{8})\right)\\
& \le \,\frac{12.8}{d_{i-1}},
\end{align*}
where we have canceled the exponential terms, establishing (\ref{newalg5}).

To establish (\ref{newalg6}), note that each $d_i$ is increasing as
a function of $d_0$, and thus each $\E[X_i^2]$ is upper bounded by
taking $d_0 \ge 4$ to be 4.  Then $\E[X_1^2] = 3.2$, $\E[X_2^2] = 1.6648$,
and the other terms can be bounded in a geometric series with a sum 
less than 0.12.

Finally, to establish (\ref{newalg7}), note that
\begin{align*}
\Pr(\hat m_i \ne m)
&\mathop{=} \Pr(|U_i|^2 \ge 1)~~~ \le \E[U^2_{i+1}]\\
&\mathop{\le}^{(a)} \frac{1.6}{d_i}\exp(-d_i^2/8) \mathop{\le}^{(b)} 
\exp(-d_i^2/8)\\
&\mathop{=}^{(c)} 1/\exp( g_i(d_0^2/8)) \mathop{\le}^{(d)} 1/g_{i+1}(2),
\end{align*}
  where we have used Lemma \ref{lem:pam1} in $(a)$,  the fact that 
$d_i \ge 4$ in $(b)$, and equation (\ref{newalg4}) in $(c)$ and $(d)$.
\end{proof}

We have now shown that, in this high SNR regime, the error
  probability decreases with time $i$  as an $i$th order exponent.
The constants  involved, such as $d_0\ge 4$ are somewhat ad hoc, and
  the details of the derivation are similarly ad hoc.  What is happening,
  as stated before, is that by using PAM centered on the receiver's current
  tentative decision,  one can achieve rapidly expanding signal point
separation with  small energy. This is the critical idea driving this
algorithm, and in essence this idea was used earlier by\footnote{However 
unlike the scheme presented above, in Zigangirov's scheme the total amount of 
energy needed for transmission is increasing  linearly with time.} 
Zigangirov \cite{Zig}

\section{A two-phase strategy}\label{sectwophase}

We now combine the Shalkwijk-Kailath (SK) scheme of Section 
\ref{secSK}  and the high SNR scheme of Section \ref{secPAM} into a 
two phase strategy. The first phase, of block length $n_1$, uses the 
SK scheme.  At time $n_1-1$, the equivalent received signal
 $Y_0 - \E[Z_0 \mid {\bf Y}_1^{n_1-1}]$, (see (\ref{geonoise})), is
used in an ML decoder to detect  the original PAM signal $X_0$ in 
the presence of additive Gaussian noise of variance $\sigma_{n_1}^2$.

Note that if we scale the equivalent received signal, $Y_0 - \E[Z_0 
\mid {\bf Y}_1^{n_1-1}]$  by a factor of $1/\sigma_{n_1}$ so as to have 
an equivalent unit variance additive noise,  we see that the distance between 
adjacent signal points in the normalized PAM is $d_{n_1-1} = 2 
\gamma_{n_1}$
where $\gamma_{n_1}$ is given in (\ref{Pe1}).  If $n_1$ is selected 
to be large enough to satisfy $d_{n_1-1} \ge 4$,  then this detection 
at time $n_1-1$ satisfies the criterion assumed at time 0 of the high 
SNR  algorithm of Section \ref{secPAM}.  In other words, the SK 
algorithm not only achieves the error probability calculated in 
Section \ref{secSK}, but also, if the block length of the SK phase 
$n_1$ is chosen to be large enough, it creates the initial condition 
for the high SNR algorithm.  That is, it  provides the receiver and 
the transmitter at time $n_1-1$ with the output of a high 
signal-to-noise ratio PAM. Consequently not only is the tentative ML 
decision at time $n_1-1$ correct with moderately high probability, 
but also the probability of the distant neighbors of the decoded 
messages vanishes rapidly.

The intuition behind this two-phase scheme is that the SK algorithm 
seems  to be quite efficient when the signal points are so close 
(relative to the noise) that the discrete nature of the signal 
is not of great benefit.  When the SK scheme is used enough 
times,  however, the signal points becomes far apart relative to the 
noise, and the discrete nature of the signal becomes important. The 
increased effective distance between the signal points of the 
original PAM also makes the high SNR scheme, feasible.
 Thus the two-phase strategy switches to 
the high SNR scheme at this point and the high SNR scheme drives the 
error probability to 0 as an $n_2$ order exponential.

We now turn to the detailed analysis of this two-phase scheme.  Note 
that  5 units of energy must be reserved for phase 2 of the 
algorithm, so the power constraint $S_1$  for the first phase of 
the algorithm is  $n_1S_1=n S-5$. For any fixed rate $R< C(S)$, we 
will find that  the remaining $n_2=n-n_1$ time units is a linearly 
increasing function of $n$ and yields an error probability upper 
bounded by $1/g_{n_2+1}(2)$.

\subsection{The finite-bandwidth case}
For the finite-bandwidth case, we assume  an overall block length $n 
= n_1+n_2$, an overall power constraint $S$, and an overall rate  $R 
= (\ln M)/n$.  The overall energy  available for phase $1$ is at 
least $nS-5$, so the average power  in phase $1$ is at least 
$(nS-5)/n_1$.

\begin{figure}[h]
\setlength{\unitlength}{5pt}
\centering
\begin{picture}(76,18)(-8,-3.5)
\put(0,0){\line(1,0){40}}
\put(30,0){\line(0,1){16}}
\put(17.3,11){\line(0,-1){11}}
\put(12.4,11){\line(0,-1){11}}
\qbezier(0,0)(0,7.21)(15,11.88)
\qbezier(30,15)(20.97,13.74)(15,11.88)
\qbezier(0,5.66)(4.3,7)(30,15)
\put(31,14.4){\footnotesize $C$}
\put(31,6.4){\footnotesize $R$}
\put(31,10.4){\footnotesize $R+\beta/n$}
\put(12,-2.1){\footnotesize $\nu'_n$}
\put(30,11){\line(-1,0){17.6}}
\put(3,-2.1){\footnotesize $\phi^{-1}(R)$}
\put(0,-2.1){\footnotesize $0$}
\put(16.8,-2.1){\footnotesize $\nu_n$}
\put(29.7,-2.1){\footnotesize $1$}
\put(22,.6){\footnotesize $\nu$}
\put(21,14.5){\footnotesize $\phi(\nu)$}
\put(30,7){\line(-1,0){25.7}}\put(4.3,7){\line(0,-1){7}}
\end{picture}\vspace{-.3cm}\caption{\label{figfofnu} \small This
shows the function $\phi(\nu)$ and also the value of $\nu$, denoted
$\phi^{-1}(R)$, at which $\phi(\nu)=R$.  It also shows $\nu_n'$,
which satisfies $\phi(\nu_n') = R+\beta/n$, and gives the solution to
(\ref{fofnu1}) with equality. It turns out to be more convenient to
satisfy (\ref{fofnu1}) with inequality using $\nu_n$, which by simple
geometry satisfies $\nu_n = \phi^{-1}(R) +
\frac{\beta(1-\phi^{-1}(R))}{n(C-R)} $. }\vspace{-0.2cm}
\end{figure}
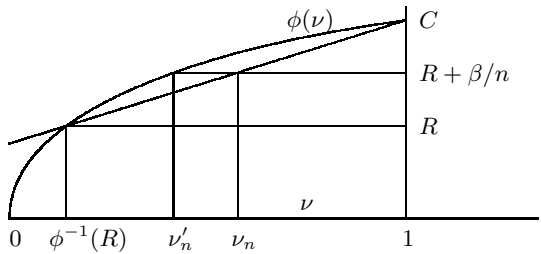

We observed that the distance $d_{n_1-1}$ between adjacent signal 
points, assuming that signal and noise are normalized to unit noise 
variance, is
twice the parameter $\gamma_{n_1}$ given in (\ref{Pe3}).  Rewriting   
(\ref{Pe3}) for the power constraint $(nS-5)/n_1$,
\begin{align}
\notag
d_{n_1}
&\ge
2\sqrt{3}\left(1+\frac{nS-5}{n_1}-\frac{1}{n_1}\right)^{n_1/2}\exp(-nR)\\
\notag
&=
2\sqrt{3}\left(1+\frac{nS}{n_1}\right)^{n_1/2}\exp(-nR)\left(1-\frac{6}{nS+n_1} 
\right)^{n_1/2}\\
\notag
&\mathop{\geq}^{(a)}
2\sqrt{3}\left(1+\frac{nS}{n_1}\right)^{n_1/2}\exp(-nR)\left(1-\frac{1}{1+n_1/6} 
\right)^{n_1/2}\\
\label{Peph2}
&\ge
\tfrac{2\sqrt{3}}{e^{3}} \left(1+\frac{Sn}{n_1}\right)^{n_1/2}\exp(-nR),
\end{align}
where to get $(a)$ we assumed that $nS\ge 6$. We can also show that 
the  multiplicative term, $(1-\tfrac{1}{1+n_1/6})^{n_1/2}$, is  a 
decreasing function of $n_1$ satisfying
\begin{equation*}
   \left(1-\frac{1}{1+n_1/6} \right)^{n_1/2} \geq \lim_{n_1 
\rightarrow \infty}\left(1-\frac{1}{1+n_1/6} \right)^{n_1/2} = 
e^{-3}.
\end{equation*}
This establishes  (\ref{Peph2}). In order to satisfy  $d_{n_1}\ge 4$, 
it suffices for the right-hand side of (\ref{Peph2}) to be greater 
than or equal to $4$.  Letting $\nu = n_1/n$, this condition can be 
rewritten as
\begin{equation}
\label{fofnu}
\exp\left[n\left(-R + \frac{\nu}{2}\ln (1 + 
\frac{S}{\nu}\right)\right] \ge  \tfrac{2 e^3}{\sqrt{{3}}}.
\end{equation}
Define $\phi(\nu)$ by
\[
\phi(\nu) = \frac{\nu}{2}\ln(1+S/\nu).
\]
This is a concave
increasing function for $0<\nu \le 1$ and can be interpreted as the
capacity of the given channel if the number of available degrees of
freedom is reduced from $n$ to $\nu n$ without changing the available
energy per block, \ie, it can be interpreted as the capacity of a
continuous time channel whose bandwidth has been reduced by a factor
of $\nu$.    We can then rewrite (\ref{fofnu}) as
\begin{equation}
\label{fofnu1}
\phi(\nu) \ge R + \frac{\beta}{n},
\end{equation}
where $\beta = \ln (\tfrac{2 e^3}{\sqrt{3}}) $.  This is interpreted in Figure
\ref{figfofnu}.

The condition $d_{n_1} \ge 4$ is satisfied by choosing $n_1 =
\lceil n \nu_n\rceil$ for $\nu_n$ defined  in Figure \ref{figfofnu},
\ie, \[ n_1 = \left\lceil n\phi^{-1}(R)  +
\frac{\beta(1-\phi^{-1}(R))}{C-R}  \right\rceil\]Thus the duration
$n_2$ of phase 2 can be chosen to be
\begin{equation}
\label{n2lin}
n_2 = \left\lfloor n[1-\phi^{-1}(R)] -
\frac{\beta(1-\phi^{-1}(R))}{C-R}\right\rfloor.
\end{equation}

This shows that  $n_2$ increases linearly with $n$ at rate 
$1-\phi^{-1}(R)$ for $n > \beta/(C-R)$. As a result of lemma 
\ref{lem:HighSNR}  the error probability is upper bounded as
\begin{equation}
\label{order3}
   \Pr(\hat m \ne m) \le 1/g_{n_2+1}(2),
\end{equation}
  Thus the probability of  error is bounded by an exponential order 
that increases at a rate $1-\phi^{-1}(R)$. We later derive a lower 
bound to error probability which has this same rate of increase for 
the  exponential order of error probability.

\subsection{The broadband case - zero error probability}
The broadband case is somewhat simpler since an unlimited number of 
degrees of  freedom are available.  For phase 1, we start with 
equation  (\ref{Pe6}), modified by the fact that $5$ units of energy 
must be reserved for phase 2.
\begin{align}
\notag
d_{n_1}
& \ge  2\sqrt{3}\exp\left[\frac{n_1}{2}\ln(1+\frac{\mathcal PT}{n_1} 
- \frac{6}{n_1})-T R_{\infty}\right]\\
\notag
&\ge  2\sqrt{3}\exp\left[\frac{\mathcal P T}{2} - 3 - \frac{\mathcal 
P^2T^2}{4n_1}-T R_\infty\right],
\end{align}
where, in order to get the inequality in the second step, we assumed 
that $\mathcal P T\geq 6$ and used the identity  $\ln(1+x) \ge 
x-x^2/2$. As in the
broadband SK analysis,  we assume that $n_1$ is increasing 
quadratically with increasing $T$. Then  $\tfrac{\mathcal 
P^2T^2}{4n_1}$ becomes just  a constant. Specifically if $n_1\geq 
\tfrac{\mathcal P^2T^2}{4}$ we get,
\begin{equation*}
d_{n_1} \ge \tfrac{ 2\sqrt{3}}{e^4}\exp\left[T(C_{\infty}-R_\infty)\right],
\end{equation*}
It follows that $d_{n_1} \ge 4$ if
\begin{equation}
\label{Pe9ph2}
T \ge \tfrac{4+ \ln 2-0.5 \ln 3}{C_\infty - R_\infty}.
\end{equation}
If (\ref{Pe9ph2}) is satisfied, then phase 2 can be carried out for
arbitrarily large $n_2$, with $P_e$ satisfying (\ref{order3}).  In
principle, $n_2$ can be infinite, so $P_e$ becomes 0 whenever $T$ is
large enough to satisfy(\ref{Pe9ph2}).

One might object that the transmitter sequence is not well defined
with $n_2=\infty$, but in fact it is, since at most a finite number
of transmitted symbols can be nonzero.  One might also object that it
is impossible to obtain an infinite number of ideal feedback signals
in finite time. This objection is certainly valid, but the entire idea of
ideal feedback with infinite bandwidth  is unrealistic.  Perhaps a more
comfortable way to express this result is that 0 is the greatest
lower bound to error probability when (\ref{Pe9ph2}) is satisfied,
\ie, any desired error probability, no matter how small is achievable
if the continuous-time  block length $T$ satisfies (\ref{Pe9ph2}).

    \section{A lower bound to error probability}
The previous sections have derived upper bounds to the probability of
decoding error for data transmission using particular block coding
schemes with ideal feedback.    These schemes  are non-optimal, with
the non-optimalities chosen both for analytical convenience and for
algorithmic simplicity.  It appears that the optimal strategy is
quite complicated and probably not very interesting.  For example,
even with a block length $n=1$, and a message set size $M=4$,  PAM
with equi-spaced messages is neither optimal in the sense of
minimizing average error probability over the message set (see
Exercise 6.3 of \cite{Gall2}) nor in the sense of minimizing the
error probability of the worst message.  Aside from this rather
unimportant non-optimality, the SK scheme is also non-optimal in
  ignoring the discrete nature of the signal until the final
decision. Finally, the improved algorithm of Section
\ref{sectwophase} is non-optimal both in
using ML rather than maximum \emph{a posteriori}
probability (MAP) for the tentative decisions  and in not
optimizing the choice of signal points as a function of the prior
received signals.

The most important open question, in light of the extraordinarily
rapid decrease of error probability with block length for the finite
  bandwidth case, is whether  any strictly positive lower bound to
error probability exists for fixed block length $n$. To demonstrate
that there is such a positive lower bound we first derive a lower
bound to error probability for the special case of a message set of
size $M=2$. Then we generalize this to codes of arbitrary rate and show
that for $R<C$, the lower bound decreases as a $k$th order 
exponential where $k$ increases with the block length $n$ and has the 
form $k = an-b'$ where the coefficient $a$ is the same as that in the
upper bound in Section \ref{sectwophase}.  It is more convenient in 
this section  to number the
successive signals from 1 to $n$ rather than $0$ to $n-1$ as in previous
sections.

\subsection{A lower bound for $M=2$}\label{binary}
Although it is difficult to find and evaluate the entire optimal
code, even for $M=2$, it turns out to be easy to find  the optimal
encoding in the last step.   Thus, for each $\vec Y_1^{n-1}$, we want 
to find the optimal choice of $X_n = f(U, \vec Y_1^{n-1})$ as a 
function of, first, the encoding functions $X_i = f(U, \vec 
Y_1^{i-1})$,  $1\le i \le n-1$, and, second,  the
allocation of energy,  $\tilde S = \E[X_{n}^2|\vec Y_1^{n-1}]$ for that $\vec
Y_1^{n-1}$.  We will evaluate the error
probability for such an optimal  encoding at time $n$ and then
relate it to the error probability that would have resulted from
decoding at time $n-1$.  We will use this relation to develop a recursive
lower bound to error probability at each time $i$ in terms of that at
time $i-1$.

  For a given code function $X_i = f(U, \vec Y_1^{i-1})$ for $1\le i
\le n-1$,  the conditional probability density\footnote{We do not
use the value of  this density, but for completeness, it can be seen
  to be   $\prod_{j=1}^i \xi[Y_j - f(U, \vec Y_1^{j-1})]$ where
$\xi(x)$ is the normal  density $(2\pi)^{-1/2}\exp(-x^2/2)$. }
  of $\vec Y_1^i$ given $U=1$ or 2 is positive for all sample values for
$\vec Y_1^i$; thus  the corresponding conditional probabilities of
hypotheses $U=1$ and $U=2$ are positive i.e.
\begin{equation*}
  \Pr(U{=} m| \vec Y_1^{i})>0 \qquad m \in \{1, 2\},~ \forall \vec Y_1^i\in \mathbb{R}^{i}.
\end{equation*}
  In particular,  for $m \in 
\{1, 2\}$, define $\Phi_m = \Pr(U{=}m | \vec Y_1^{n-1})$  for some 
given
$\vec Y_1^{n-1}$. Finding the error probability $\Psi = 
\Pr(\hat{U}(\vec Y_1^{n})\neq U \mid  \vec Y_1^{n-1})$   is an
elementary binary detection problem for the given
$\vec Y_1^{n-1}$.  \,MAP detection, using the \emph{a
priori} probabilities $\Phi_1$ and $\Phi_2$,  minimizes the resulting
error probability.

For a given sample value of $\vec Y_1^{n-1}$, let $b_1$ and $b_2$ be 
the values of $X_n$  for
$U = 1$ and $2$ respectively.  Let $a$ be half the distance between
$b_1$ and $b_2$, \ie, $2a = b_2-b_1$.  The error probability $\Psi$
depends on $b_1$
and $b_2$ only through $a$.  For a given $\tilde S$, we choose $b_1$ and $b_2$
to satisfy $\E[X_{n}|\vec Y_1^{n-1}] = 0$, thus
maximizing $a$ for the given $\tilde S$. The variance of $X_n$
conditional on $\vec Y_1^{n-1}$ is given by $$\mbox{Var}(X_n | \vec
Y_1^{n-1}) =  \frac{1}{2}\sum_{i, j}\Phi_i\Phi_j(b_i-b_j)^2 =
4\Phi_1\Phi_2a^2,$$and since $\E[X_{n}|\vec Y_1^{n-1}] = 0$, this
means that $a$ is related to $\tilde S$ by  $\tilde S = 4\Phi_1\Phi_2a^2$.

Now let $\Phi = \min\{\Phi_1, \Phi_2\}$. Note that $\Phi$ is the
probability of error for  a hypothetical MAP decoder detecting $U$
at time $n-1$ from $\vec Y_1^{n-1}$. The error probability $\Psi$ for
the MAP decoder at the end of time $n$ is  given by the classic result
of binary MAP detection with \emph{a priori} probabilities $\Phi$ and
$1-\Phi$,
   \begin{equation}
\label{exact}
\Psi = (1-\Phi)Q\left(a + \frac{\ln \eta}{2a}\right) + \Phi Q\left(a
- \frac{\ln \eta}{2a}\right),
\end{equation}
where $\eta = \tfrac{1-\Phi}{\Phi}$ and $Q(x) = \int_x^\infty 
(2\pi)^{-1/2} \exp(-z^2/2)\,dz$.  This equation relates the error 
probability
$\Psi$ at the end of time $n$ to the error probability $\Phi$ at the
end of time $n-1$, both conditional on  $\vec Y_1^{n-1}$.  We are now
going to view $\Psi$ and $\Phi$ as functions of $\vec Y_1^{n-1}$, and
thus as random variables. Similarly  $\tilde S\ge 0$ can be any
non-negative  function of
$\vec Y_1^{n-1}$,   subject to a  constraint $S_n$ on its mean;
so we can view $\tilde S$ as an arbitrary non-negative random variable
with mean $S_n$.  For each $\vec Y_1^{n-1}$, $\tilde{S}$ and $\Phi$
determine the value of $a$; thus $a$ is also a non-negative
random variable.

  We are now going to  lower bound the expected value of $\Psi$ in such
a way that the  result is a function only  of the expected value of
  $\Phi$ and the  expected value $S_n$ of $\tilde S$.  Note that $\Psi$ in
(\ref{exact}) can be lower bounded by ignoring the first term and
replacing the second term with $\Phi Q(a)$.    Thus,

\begin{align}
\Psi 
\notag
&\ge \Phi Q(a)\\
\notag
&= \Phi Q\left(\sqrt{\frac{\tilde S}{4\Phi (1-\Phi
)}}\right) \\
&\ge  \Phi Q\left(\sqrt{\frac{\tilde S}{2\Phi  }}\right) .
\label{lb}
\end{align}
where the  last step uses the facts that $Q(x)$ is a decreasing function
of $x$ and that $1-\Phi > 1/2$.

\begin{eqnarray}
\label{ineq1}
\E[\Psi ]
&\ge&
\E[\Phi ] Q\left(\frac{1}{\E[\Phi ]}\E\left[\Phi \sqrt{\frac{\tilde
S}{2\Phi  }}\right]\right)\\
&=&\nonumber
\E[\Phi ] Q\left(\frac{1}{\sqrt{2}\E[\Phi ]}\E\left[\sqrt{\Phi \tilde
S}\right]\right)\\
\label{ineq2}
&\ge&
\E[\Phi ] Q\left(\frac{1}{\sqrt{2}\E[\Phi ]}\sqrt{\E[\Phi ]\E [\tilde
S]}\right)\\
&=&\E[\Phi ] Q\left(\sqrt{\frac{S_n}{2\E[\Phi ]}}\,\right).
\label{ineq3}
\end{eqnarray}
In (\ref{ineq1}), we used Jensen's inequality, based on the facts
that  $Q(x)$ is a convex function for $x\ge 0$ and that $\Phi
/\E[\Phi ]$ is a probability distribution on $\vec Y_1^{n-1}$.  In
(\ref{ineq2}), we used the Schwarz inequality along with the fact
that $Q(x)$ is decreasing for $x\ge 0$.

We now recognize that $\E[\Psi]$ is simply the overall error
probability at the end of time $n$ and $\E[\Phi]$ is the overall
error probability (if a MAP decision were made) at the end of time
$n-1$. Thus we denote these quantities as $p_n$ and $p_{n-1}$
respectively,
\begin{equation}
\label{lagrange6}
p_{n} \ge p_{n-1}Q\left(\sqrt{\frac{S_{n}}{2p_{n-1}}}\right).
\end{equation}
Note that this lower bound is monotone increasing in $p_{n-1}$.  Thus
we can further lower bound $p_n$ by lower bounding $p_{n-1}$.  We can
lower bound $p_{n-1}$ (for a given $p_{n-2}$ and $S_{n-1}$) in exactly
the same way, so that $p_{n-1} \ge 
p_{n-2}Q(\sqrt{S_{n-1}/2p_{n-2}})$.    These two bounds can  be
combined to implicitly bound $p_{n}$ in terms of $p_{n-2}$, $S_n$ and
$S_{n-1}$.  In fact, the same technique can be used for each $i, 1\le
i \le n$, getting
\begin{equation}
\label{lagrange5}
p_i \ge p_{i-1}Q\left(\sqrt{\frac{S_i}{2p_{i-1}}}\right).
\end{equation}
This gives us a recursive lower bound on $p_{n}$ for any given choice
of $S_1, \ldots, S_{n}$ subject to the power constraint $\sum_i S_i
\le nS$.

We have been unable to find a clean way to optimize this over the
choice of $S_1, \ldots, S_n$, so as a very crude lower bound on
$p_n$, we upper bound each $S_i$ by $nS$. For convenience, multiply
each side of  (\ref{lagrange5}) by $2/nS$,
\begin{equation}
\label{lagrange7}
\frac{2p_i}{nS} \ge \frac{2 p_{i-1}}{nS} Q\left(\sqrt{\tfrac{nS}{2
p_{i-1}}}\right);\qquad \quad \mbox{for} \,\,1\le i \le n.
\end{equation}
   At this point, we can see what is happening  in this lower bound.
As $p_i$ approaches 0,~ $\tfrac{nS}{2p_i} \to \infty$.  Also
$Q\left(\sqrt{\tfrac{nS}{2p_i}} \right)$ approaches $0$ as
$e^{-\frac{nS}{4p_i}}$. Now we will lower bound the expression on the
right hand side of (\ref{lagrange7}). We can check
numerically\footnote{That is, we can check numerically that 
(\ref{ninebound}) is satisfied for $x=9$ and verify that the 
right-hand side is decreasing faster than the left for $x>9$. } that 
for $x\ge 9$,
\begin{equation}\label{ninebound}
\frac{1}{x} Q(\sqrt{x})  \ge  \exp( -x).
\end{equation}
Furthermore $\frac{1}{x} Q(\sqrt{x})$ is decreasing in $x$ for all 
$x>0$, and thus
\begin{equation*}
\frac{1}{x} Q(\sqrt{x})  \ge  \exp( -\max\{x, 9\}) \qquad \forall x>0.
\end{equation*}
Substituting this into (\ref{lagrange7}) we get,
\begin{equation*}
\frac{2p_i}{nS} \ge \frac{1}{\exp( \max\{ \frac{nS}{2 p_{i-1}},9
\})};\qquad \quad \mbox{for} \,\,1\le i \le n.
\end{equation*}
Applying this recursively for $i = n$ down to $i=k+1$ for any $k\ge 0$ we get,
\begin{align}
\frac{2p_n}{nS}
\notag
&\ge \frac{1}{\exp( \max\{\exp (\max \{\frac{nS}{2 p_{n-2}},9 \}) ,9\})}\\
\notag
&\mathop{=}^{(a)} \frac{1}{\exp(\exp (\max \{\frac{nS}{2 p_{n-2}},9 \}))}\\
\label{nkorder}
&\ge \frac{1}{g_{n-k}\left[\max\left\{\tfrac{nS}{2p_{k}}, 9\right\}\right]}.
\end{align}
where $(a)$ simply follows from the fact that $\exp(9)>9$. This bound
holds for $k=0$, giving an overall lower bound on error probability
in terms of $p_0$.  In the usual case where the symbols are initially
equiprobable, $p_0 = 1/2$ and
\begin{equation}
\label{xn}
p_n \ge \frac{nS}{2g_n[\max(nS,9)]}.
\end{equation}
Note that this lower bound is an  $n$th order exponential. Although 
it is numerically much smaller than the upper bound in Section 
\ref{sectwophase}, it has the same general form.  The
intuitive interpretation is also similar.  In going
from block length $n-1$ to $n$, with very small error
probability at $n-1$, the symbol of large \emph{a priori} probability is
very close to 0 and the other symbol is approximately at 
$\sqrt{\tilde S/p_{n-1}}$.  Thus the error probability is decreased 
in one time unit by an exponential in $p_{n-1}$, leading to an $n$th order
 exponential over $n$ time units.

\subsection{Lower bound for arbitrary $M$}
Next consider feedback codes of arbitrary rate $R<C$ with
sufficiently large blocklength $n$ and $M = e^{nR}$ codewords.  We
derive a lower bound on error probability  by splitting $n$ into an
initial segment of length $n_1$ and a final segment of length $n_2 =
n-n_1$.  This segmentation is for bounding purposes only and does not
restrict the feedback code.  The error probability of a hypothetical
MAP decoder at the end of the first segment,  $P_e(n_1)$, can be
lower bounded by a conventional use of the Fano inequality. We will
show how to use this error probability as the input of the lower
bound for $M=2$ case derived  in the previous subsection, i.e.,
equation (\ref{nkorder}).  There is still the question of allocating
power between the two segments, and since we are deriving a lower
bound, we simply assume that the entire available energy is available
in the first segment, and can be reused in the second segment. We
 will find that the resulting lower bound has the same form as the
upper bound in Section \ref{sectwophase}.

Using energy $Sn$ over the first segment corresponds to power
$Sn/n_1$, and since feedback does not increase the channel capacity,
the average directed mutual information over the first segment is at
most $n_1C(Sn/n_1)$. Reusing the definitions $\nu = n_1/n$ and
$\phi(\nu) = \frac{\nu}{2}\ln(1+\frac{S}{\nu})$ from Section
\ref{sectwophase}, 
\[n_1C(Sn/n_1) = n\phi(\nu).\]
The entropy of the
source is $\ln M = nR$, and thus the conditional entropy of the
source given $\vec Y_1^{n_1}$   satisfies
\begin{eqnarray}
\notag
%\label{condent}
n\left[R - \phi(\nu) \right]  & \le& H(U | \vec Y_1^{n_1}) \\
\notag
%\label{Fanoineq}
&\le& h(P_e(n_1)) + P_e(n_1)nR\\
&\le& \ln 2 + P_e(n_1)nR,
\label{Fano2}
\end{eqnarray}
where we have used the Fano inequality and then bounded the binary
entropy $h(p) = -p \ln p -(1-p)\ln (1-p)$ by $\ln 2$.

\begin{figure}[h]
\setlength{\unitlength}{5pt}
\centering
\begin{picture}(76,18)(-8,-2)
\put(0,0){\line(1,0){40}}
\put(30,0){\line(0,1){16}}
\put(30,8.5){\line(-1,0){26.1}}
\put(3.9,8.5){\line(0,-1){8.5}}
\put(6.75,8.5){\line(0,-1){8.5}}
\put(30,10){\line(-1,0){20.2}}
\put(9.8,10){\line(0,-1){10}}
\qbezier(0,0)(0,7.21)(15,11.88)
\qbezier(30,15)(20.97,13.74)(15,11.88)
\qbezier(0,7.575)(9.8,10)(30,15)
\put(31,14.4){\footnotesize $C$}
\put(31,9.7){\footnotesize $R$}
\put(31,7.6){\footnotesize $R-1/n$}
\put(6,-2.1){\footnotesize $\nu'_n$}
\put(9,-2.1){\footnotesize $\phi^{-1}(R)$}
\put(0,-2.1){\footnotesize $0$}
\put(3,-2.1){\footnotesize $\nu_n$}
\put(29.7,-2.1){\footnotesize $1$}
\put(20,.6){\footnotesize $\nu$}
\put(18,14){\footnotesize $\phi(\nu)$}
\end{picture}
\vspace{0cm}
\caption{\footnotesize This shows the  value of $\nu$, denoted
$\phi^{-1}(R)$, at which $\phi(\nu)=R$.  It also shows $\nu_n'$,
where $\phi(\nu_n') = R-1/n$.  This gives the solution to
(\ref{boundn1}) with equality, but $\nu_n = \phi^{-1}(R) -
\frac{1-\phi^{-1}(R)}{n(C-R)}$ can be seen to be less than $\nu_n'$
and thus also satisfies (\ref{boundn1}).}
\vspace{-0.5cm}
\label{figfofnu1}
\end{figure}
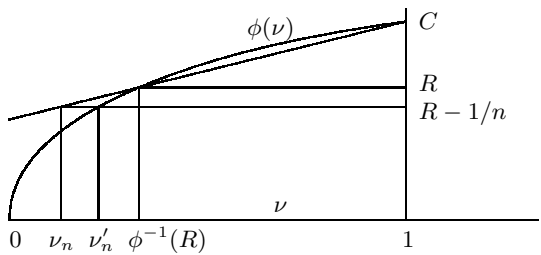

To use (\ref{Fano2}) as a lower bound on $P_e(n_1)$, it is necessary
for $n_1 =n\nu$ to be small enough that $\phi(\nu)$  is substantially
less than $R$, and to be specific we choose $\nu$ to satisfy
\begin{equation}
\label{boundn1}
R-\phi(\nu) \ge \frac{1}{n}.
\end{equation}
With this restriction, it can be seen from (\ref{Fano2}) that
\begin{equation}
\label{Fano}
P_e(n_1) \ge \frac{1 - \ln 2}{nR}.
\end{equation}
Figure \ref{figfofnu1} illustrates that the following choice of 
$n_1$ in (\ref{boundn1a}) satisfies both equation (\ref{boundn1}) and 
equation (\ref{Fano}).
This uses the fact that $\phi(\nu)$ is a monotonically increasing 
concave function of $\nu$.
\begin{equation}
\label{boundn1a}
n_1 = \left\lfloor n\phi^{-1}(R) - \frac{1-\phi^{-1}(R)}{C-R} \right\rfloor.
\end{equation}

The corresponding choice for $n_2$ is
\begin{equation}
\label{boundn2a}
n_2 = \left\lceil n[1-\phi^{-1}(R)] + \frac{1-\phi^{-1}(R)}{C-R} \right\rceil.
\end{equation}
Thus with this choice of $n_1, n_2$, the error probability at the end
of time $n_1$ satisfies (\ref{Fano}).

The straightforward approach at this point would be to generalize the
recursive relationship in (\ref{lagrange5}) to arbitrary $M$.  This
recursive relationship could then be used, starting at time $i=n$ and
using each successively smaller $i$ until terminating the recursion
at $i = n_1$ where (\ref{Fano}) can
be used.  It is simpler, however, since we have already derived
(\ref{lagrange5}) for $M=2$, to define a binary coding scheme from
any given $M$-ary scheme in such a way that the binary results can be
used to lower bound the $M$-ary results. This technique is similar to
one used earlier in \cite{Telatar}.

   Let $X_i = f(U, \vec Y_1^{i-1})$ for $1\le i \le n$ be any given
coding function for $U\in \mathcal M = \{1, \ldots, M\}$. That
code is used to define a related binary code.  In particular, for  each
received sequence $\vec Y_1^{n_1}$ over the first segment, we
partition the message set $\cal M$ into two subsets, $\mathcal
M_1(\vec Y_1^{n_1})$ and $\mathcal M_2(\vec Y_1^{n_1})$. The
particular partition for each $\vec Y_1^{n_1}$ is defined later.
This partitioning defines a binary random variable $V$  as follows,
   \begin{equation*}
V=
\begin{Bmatrix}
   1&~&U\in \mathcal M_1(\vec Y_1^{n_1})\\
   2&~&U\in \mathcal M_2(\vec Y_1^{n_1})
\end{Bmatrix}
   \end{equation*}
At the end of the transmission, the receiver will use its decoder to
decide $\hat{U}$. We define the decoder for $V$ at time $n$, using
the decoder of $U$ as follows,
\begin{equation*}
\hat{V}=
\begin{Bmatrix}
   1&~&\hat{U}\in \mathcal M_1(\vec Y_1^{n_1})\\
   2&~&\hat{U}\in \mathcal M_2(\vec Y_1^{n_1})
\end{Bmatrix}
   \end{equation*}
Note that with the above mentioned definitions, whenever the $M$-ary
scheme decodes correctly, the related binary scheme does also, and
thus the error probability $P_e(n)$ for the $M$-ary scheme must be
greater than or equal to the error probability $p_n$ of the related
binary scheme.

The binary scheme, however, is one way (perhaps somewhat bizarre) of
transmitting a binary symbol, and thus it satisfies the
results\footnote{This is not quite as obvious as it sounds.  The
binary scheme here is not characterized by a coding function $f(V,
\vec Y_1^{i-1})$ as in Section \ref{binary} , but rather is a
randomized binary scheme.  That is, for a given $\vec Y_1^{n_1}$ and
a given choice of $V$, the subsequent transmitted symbols $X_i$ are
functions not only of $V$ and $\vec Y_1^{i-1}$, but also of a random
choice of $U$ conditional on $V$.  The basic conclusion of
(\ref{lagrange5}) is then justified by averaging over both $\vec
Y_1^{i-1}$ and the choice of $U$ conditional on  $V$.} of section
\ref{binary}.  In particular, for the binary scheme, the error
probability $p_n$ at time $n$ is lower bounded by the error
probability $p_{n_1}$ at time $n_1$ by (\ref{nkorder}),
\begin{equation}
\label{nkorderb}
P_e(n) \ge p_n \ge
\tfrac{nS}{2} \tfrac{1}{g_{n_2}\left[\max\left\{\tfrac{nS}{2p_{n_1}}, 
9\right\}\right]}.
\end{equation}
   Our final task is to relate the error probability $p_{n_1}$ at time
$n_1$ for the binary scheme to  the error probability $P_e(n_1)$ in
(\ref{Fano}) for the $M$-ary scheme.  In order to do this, let
$\Phi_m(\vec Y_1^{n_1})$ be the probability of message $m$
conditional on the received first segment $\vec Y_1^{n_1}$.  The MAP
error probability for an $M$-ary decision at time $n_1$, conditional
on $\vec Y_1^{n_1}$, is $1- \Phi_{\max}(\vec Y_1^{n_1})$ where
$\Phi_{\max}(\vec Y_1^{n_1}) = \max\{\Phi_1(\vec Y_1^{n_1}), \ldots
\Phi_M(\vec Y_1^{n_1})\}$.  Thus $P_e(n_1)$, given in (\ref{Fano}),
is the mean of $1-\Phi_{\max}(\vec Y_1^{n_1})$ over $\vec Y_1^{n_1}$.

   Now $p_{n_1}$ is the mean, over $\vec Y_1^{n_1}$, of the error
probability of a hypothetical MAP decoder for $V$ at time $n_1$
conditional on  $\vec Y_1^{n_1}$, $p_{n_1}(\vec Y_1^{n_1})$.  This is
the smaller of the a posteriori probabilities of the subsets
$\mathcal M_1$, $\mathcal M_2$ conditional on   $\vec Y_1^{n_1}$,
\ie,
\begin{equation}
\label{Mto2}
p_{n_1}(\vec Y_1^{n_1}) = \min\left\{
\hspace{-.4cm}\sum_{\hspace{.4cm}m\in \mathcal M_1(\vec Y_1^{n_1})} \hspace{-.7cm}\Phi_m(\vec Y_1^{n_1}), 
\hspace{-.4cm}\sum_{\hspace{.4cm}m\in \mathcal M_2(\vec Y_1^{n_1})} \hspace{-.7cm}\Phi_m(\vec Y_1^{n_1})\right\}
\end{equation}
The following lemma shows that by an appropriate choice of partition
for each $\vec Y_1^{n_1}$,  this binary error probability is lower
bounded by 1/2 the corresponding $M$-ary error probability.
   \begin{lemma}
\label{lem:con}
For any probability distribution $\Phi_1, \ldots, \Phi_M$ on a
message set ${\cal M}$ with $M>2$, let $\Phi_{\max} = \max\{\Phi_1,
\ldots, \Phi_M\}$. Then there is a partition of $\cal M$ into two
subsets, ${\cal M}_1$ and ${\cal M}_2$ such that
\begin{equation}
\label{Msplit}
\sum_{m \in {\cal M}_1} \Phi_m  \geq \frac{1-\Phi_{\max}}{2}~~
\mbox{and} ~~ \sum_{m \in {\cal M}_2} \Phi_m \geq
\frac{1-\Phi_{\max} }{2}.
\end{equation}
\end{lemma}

\begin{proof}

   Order the messages in order of decreasing $\Phi_m$. Assign the
messages one by one in this order to the sets ${\cal M}_1$ and ${\cal
M}_2$. When assigning the $k$th most likely message, we calculate the
total probability of the messages that have already been assigned to each
set, and assign the $k$th message to the set which has the smaller
  probability mass. If the probability mass of the sets are the same
we choose one of the sets arbitrarily. With such a procedure, the
difference in the probabilities of the sets, as they evolve, never
exceeds $\Phi_{\max}$.  After all messages have been assigned, let
\[ \Phi'_1 = \sum_{m\in \mathcal M_1} \Phi_m\,; \qquad \quad \Phi'_2
= \sum_{m\in \mathcal M_2} \Phi_m. \]
We have seen that $|\Phi'_1 - \Phi'_2| \le \Phi_{\max}$.  Since
$\Phi_1' + \Phi_2' = 1$, (\ref{Msplit}) follows.
\end{proof}

Since the error probability for the binary scheme is now at least one
half of that for the $M$-ary scheme for each $\vec Y_1^{n_1}$, we can
take the mean over  $\vec Y_1^{n_1}$, getting $p_{n_1} \ge
P_e(n_1)/2$.  Combining this with (\ref{nkorderb}) and (\ref{Fano})
\begin{equation}
\label{final1}
   P_e(n) \ge \frac{nS}{2}
\tfrac{1}{g_{n_2}\left[\max\left(\frac{n^2SR}{1-\ln 2},\, 9\right)\right]},
\end{equation}
where $n_2$ is given in (\ref{boundn2a}). The exact terms in this
expression are not particularly interesting because of the very weak
bounds on energy at each channel use.  What is interesting is that
the order of exponent in both the upper bound of (\ref{order3}) and
(\ref{n2lin}) and the lower bound here are increasing 
linearly\footnote{Note that the argument of $g_{n_2}$ is proportional 
to $n^2$, so that this bound  does not quite decrease with the 
exponential order $n_2$.  It does, however, decrease with an 
exponential order $n_2 +\alpha(n)$, where $\alpha(n)$ increases with 
$n$ much more slowly than, say, $\ln(\ln(n))$.  Thus $(n_2 + 
\alpha(n))/n$ is asymptotically proportional to $1-\phi^{-1}(R)$.  } 
at the same rate $1-\phi^{-1}(R)$.

\section{Conclusions}
The SK data transmission scheme can be viewed as ordinary PAM
combined with the Elias scheme for noise reduction.  The SK scheme
can also be improved by incorporating the PAM structure into the
transmission of the error in the receiver's estimate of the message,
particularly during the latter stages.  For the bandlimited version,
this leads to an error probability that decreases with an exponential
order $an+b$ where $a = 1-\phi^{-1}(R)$ and $b$ is a constant. 
 In the broadband  version, the error probability is
zero for sufficiently large finite constraint durations $T$.  A lower
bound to error probability, valid for all  $R<C$ was derived.  This
lower bound also decreases with an exponential order $an+b'(n)$ where 
again  $a = 1-\phi^{-1}(R)$ and $b'(n)$ is essentially a constant.\footnote{
$b'(n)$  is a sublinear function of $n$, i.e. 
$\displaystyle{\lim_{n \rightarrow \infty} \tfrac{b'(n)}{n}=0}$.}
 It is interesting to observe
that the strategy yielding the upper bound uses almost all the
available energy in the first phase, using at most 5 units of energy
in the second phase.  The lower bound relaxed the energy constraint,
allowing all the allowable energy to be used in the first phase and
then to be used repeatedly in each time unit of the second phase.
The fact that both bounds decrease with the same exponential order
suggests that the energy available for the second
phase is not of primary importance. An open theoretical question  is
the minimum overall energy under which the error probability for two
code words can be  zero in the infinite bandwidth case.


\begin{thebibliography}{99}
\bibitem{Telatar}
P.~Berlin, B.~Nakibo\u{g}lu, B.~Rimoldi, and E.~Telatar.
\newblock A simple converse of Burnashev's reliability function.
\newblock {\em Information Theory, IEEE Transactions on}, 
55(7):3074--3080,   July 2009.


\bibitem{Burn1}
M.~V. Burnashev.
\newblock Sequential discrimination of hypotheses with control of observations.
\newblock {\em Mathematics of the USSR-Izvestiya}, 15(3):419--440, 1980.

\bibitem{Burn}
M.~V. Burnashev.
\newblock Data transmission over a discrete channel with feedback and 
random   transmission time.
\newblock {\em Problemy Peridachi Informatsii}, 12(4):10--30, 1976.

\bibitem{Elias}
P.~Elias.
\newblock `Channel capacity without coding.
\newblock Quarterly progress report, MIT Research Laboratory of 
Electronics, Oct   15 1956.
\newblock also in \emph{Lectures on Communication System Theory}, 
E. Baghdady,  Ed., New York:McGraw Hill, 1961.


\bibitem{Gall}
R. G. Gallager,
\emph{Information Theory and Reliable Communication},  New York: Wiley, 1968.


\bibitem{Gall2}
R.~G. Gallager.
\newblock {\em Principles of Digital Communication}.
\newblock Cambridge Press, New York, 2008.

\bibitem{Kim}
Y-H. Kim, A.~Lapidoth, and T.~Weissman.
\newblock The Gaussian channel with noisy feedback.
\newblock In {\em Information Theory, 2007. ISIT 2007. IEEE 
International   Symposium on}, pages 1416--1420, June 2007.

\bibitem{Kramer}
A.~Kramer.
\newblock Improving communication reliability by use of an 
intermittent   feedback channel.
\newblock {\em Information Theory, IEEE Transactions on}, 
15(1):52--60, Jan   1969.


\bibitem{NG}
B.~Nakibo\u{g}lu and R.G. Gallager.
\newblock Error exponents for variable-length block codes with 
feedback and   cost constraints.
\newblock {\em Information Theory, IEEE Transactions on}, 
54(3):945--963, March   2008.


\bibitem{Pin}
M.~S. Pinsker.
\newblock The probability of error in block transmission in a 
memoryless  Gaussian channel with feedback.
\newblock {\em Problemy Peridachi Informatsii}, 4(4):1--14, 1968.

\bibitem{Sahai}
A.~Sahai.
\newblock Why do block length and delay behave differently if 
feedback is   present?
\newblock {\em Information Theory, IEEE Transactions on}, 
54(5):1860--1886, May   2008.



\bibitem{Schalk}
J.~Schalkwijk.
\newblock A coding scheme for additive noise channels with 
feedback--ii:  Band-limited signals.
\newblock {\em Information Theory, IEEE Transactions on}, 
12(2):183--189, Apr   1966.

\bibitem{SchalkK}
J.~Schalkwijk and T.~Kailath.
\newblock A coding scheme for additive noise channels with 
feedback--i: No   bandwidth constraint.
\newblock {\em Information Theory, IEEE Transactions on}, 
12(2):172--182, Apr   1966.

\bibitem{Shan}
C.~E. Shannon.
\newblock Two-way communication channels.
\newblock In {\em Proc. Fourth Berkeley Symp. on Math. Statist. and Prob.},
   volume~1, pages 611--644, Berkeley CA, 1961. University of California Press.

\bibitem{Zig}
K.~Sh. Zigangirov.
\newblock Upper bounds for the error probability for channels with feedback.
\newblock {\em Problemy Peredaci Informatsii}, 6(2):87--92, 1970.


\bibitem{Ozarow}
L.~Ozarow.
\newblock The capacity of the white Gaussian multiple access channel with
   feedback.
\newblock {\em Information Theory, IEEE Transactions on}, 30(4):623--629, Jul
   1984.

\bibitem{Ozarow2}
L.~Ozarow and S.~Leung-Yan-Cheong.
\newblock An achievable region and outer bound for the Gaussian broadcast
   channel with feedback (corresp.).
\newblock {\em Information Theory, IEEE Transactions on}, 
30(4):667--671, Jul   1984.


\bibitem{KramerJ}
G.~Kramer.
\newblock Feedback strategies for white Gaussian interference networks.
\newblock {\em Information Theory, IEEE Transactions on}, 
48(6):1423--1438, Jun   2002.


\bibitem{Bross}
S.I. Bross and M.A. Wigger.
\newblock On the relay channel with receiver transmitter feedback.
\newblock {\em Information Theory, IEEE Transactions on}, 
55(1):275--291, Jan.  2009.



\bibitem{Stark}
A.~Sahai, S.C. Draper, and M.~Gastpar.
\newblock Boosting reliability over awgn networks with average power 
constraints and noiseless feedback.
\newblock In {\em Information Theory, 2005. ISIT 2005. Proceedings. 
International Symposium on}, pages 402--406, Sept. 2005.


\bibitem{Deniz}
D.~G\"und\"uz, D.R. Brown, and H.V. Poor.
\newblock Secret communication with feedback.
\newblock In {\em Information Theory and Its Applications, 2008. 
ISITA 2008.  International Symposium on}, pages 1--6, Dec. 2008.

\bibitem{YHK}
Y-H. Kim.
\newblock Feedback capacity of the first-order moving average Gaussian channel.
\newblock {\em Information Theory, IEEE Transactions on}, 
52(7):3063--3079,  July 2006.

\bibitem{KimT}
Y-H. Kim.
\newblock {\em Gaussian Feedback Capacity}.
\newblock PhD thesis, Stanford University, 2006.

\end{thebibliography}
\end{document}